\documentclass[twocolumn,amsmath,amssymb]{snp}
\usepackage{graphicx}% Include figure files
\usepackage{dcolumn}% Align table columns on decimal point
\usepackage{bm}% bold math
\begin{document}

\title{{\Large Dynamics of Quark Gluon Plasma and Interference of Thermal Photons}}% 
\author{\large Dinesh K. Srivastava$^1$}
\email{dinesh@veccal.ernet.in}
\author{\large Rupa Chatterjee$^{1,2}$}
\author{\large Somnath De$^1$}

\affiliation{$^1$Variable Energy cyclotron Centre, 1/AF, Bidhan Nagar, Kolkata-700064, India}
\affiliation{$^2$Department of Physics, University of Jyv\"askul\"a,  PO Box 35 (YFL), FI-40014, Finland}

\begin{abstract}
\leftskip1.0cm
\rightskip1.0cm
The quantum statistical interference between identical particles emitted from a completely chaotic source is expected to provide valuable input for the space time description of the system. Intensity interferometry of thermal photons produced in heavy ion collisions is a very promising tool to explore the structure and dynamics of the collision fireball. Thermal photons having $K_T \, \le \, 2$ GeV/$c$ get competing contribution from both hadronic and quark matter phases and this competition gives rise to a rich structure in the outward correlation function, owing to the interference between the photons from the two sources. The temporal separation between the two sources provides the lifetime of the system and the correlation results are found to be sensitive to quark hadron phase transition temperature and the formation time of the plasma. The outward correlation function strongly depends on the equation of state of the strongly interacting matter and is seen to clearly distinguish between the lattice based and bag model equations of state. 
\end{abstract}\maketitle

\section{Introduction} 
\begin{figure}
\includegraphics[scale=0.60]{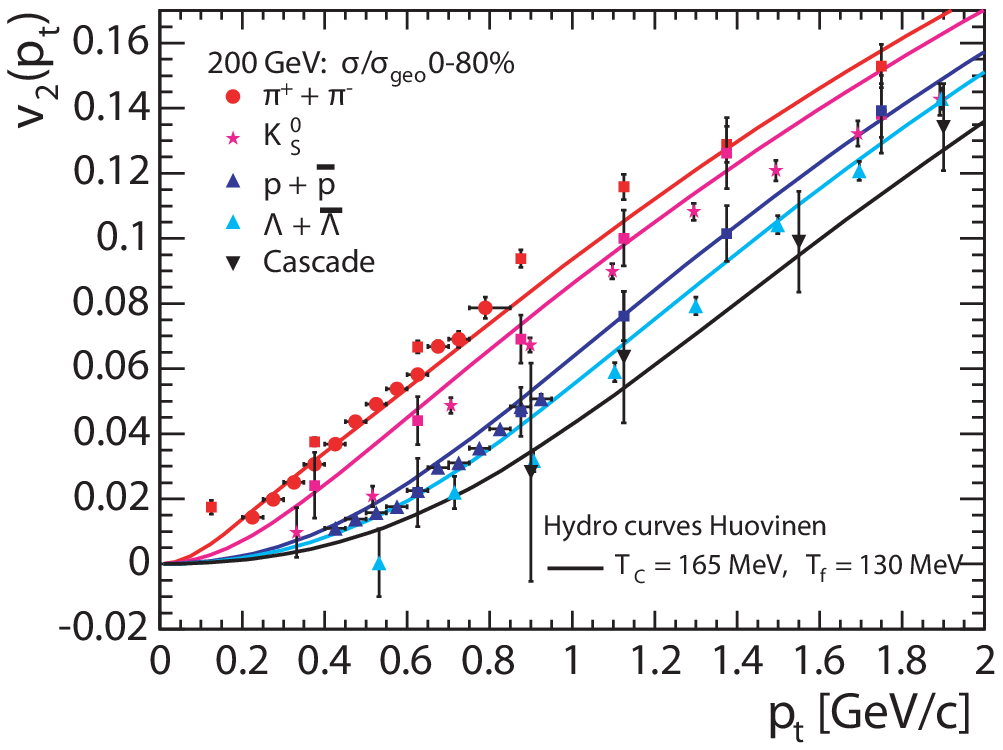}
\includegraphics[scale=0.30]{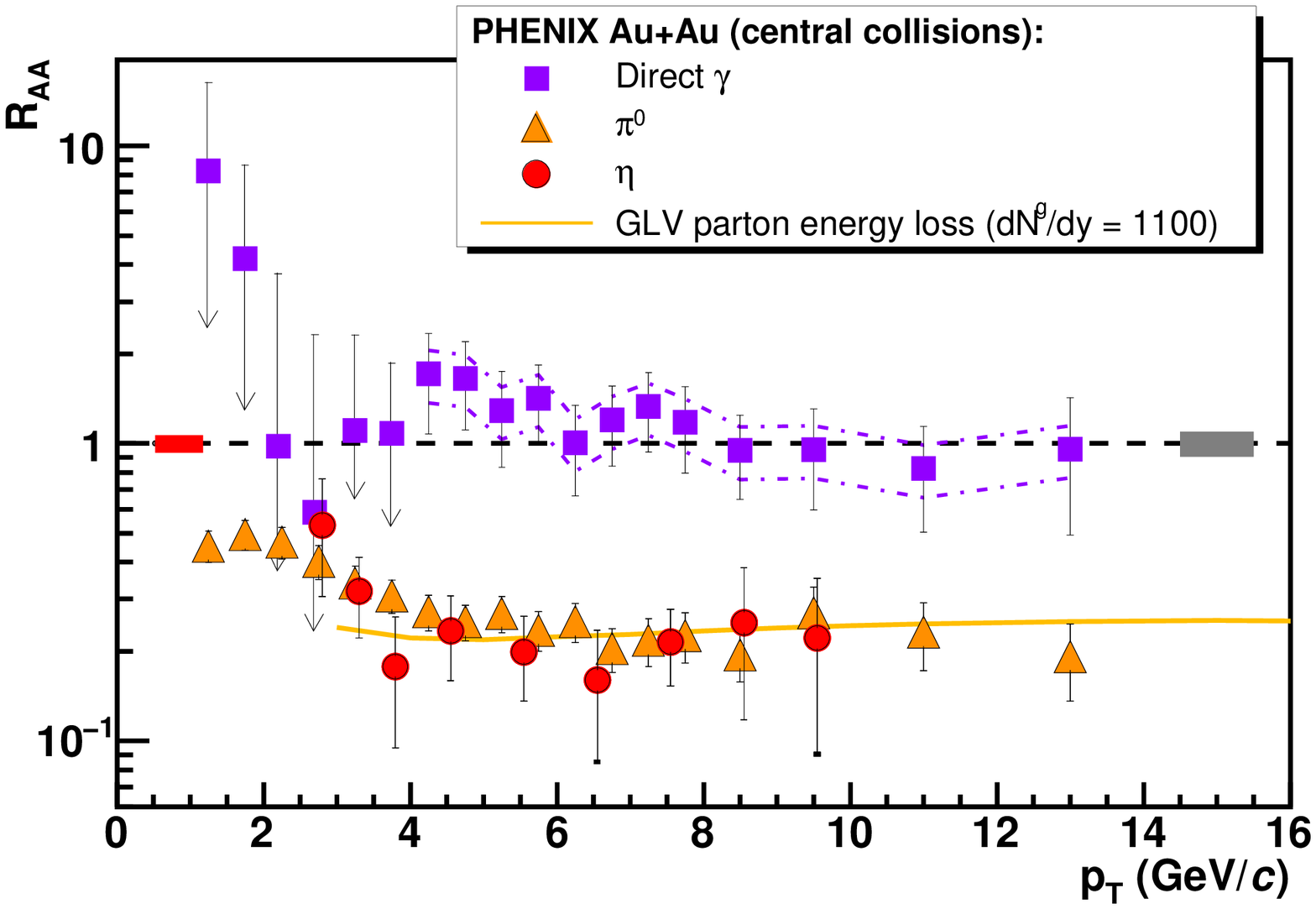}
\includegraphics[scale=0.33]{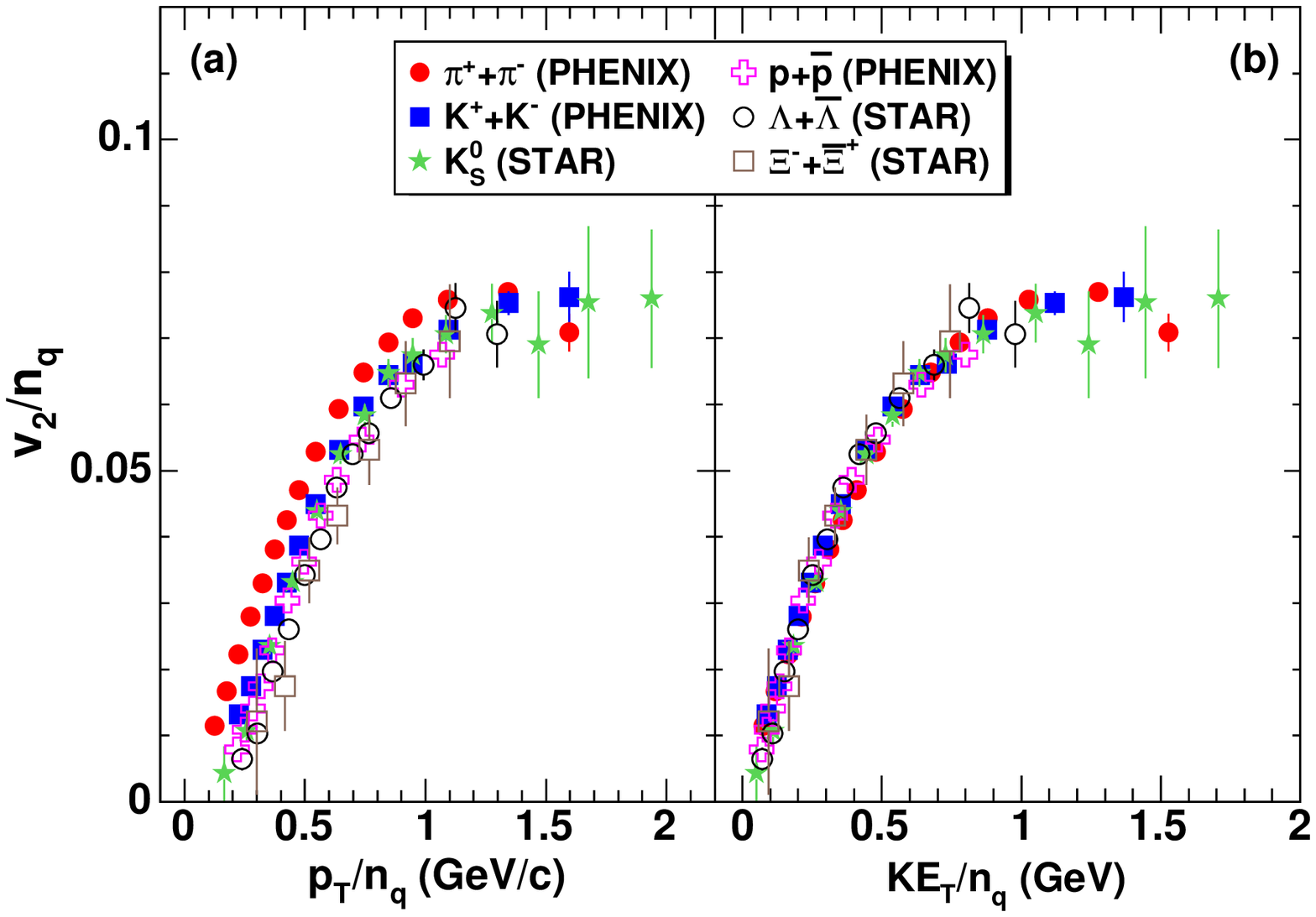}
\caption{\label{fig0.1} Elliptic flow (upper panel)~\cite{flf}, jet quenching (middle panel)~\cite{jetf}, and the success of parton recombination as a model for hadronization (lower panel)~\cite{recof} at the RHIC have advanced our understanding of the hot and dense medium produced in relativistic heavy ion collisions.}
\end{figure}

The primary motivation of the relativistic heavy ion collision studies is to investigate the properties of a deconfined strongly interacting matter, Quark Gluon Plasma (QGP) produced in the collisions.  The observation of elliptic flow~\cite{fl1,fl2}, jet quenching~\cite{jet1,jet2}, and the success of parton recombination~\cite{rec} as a model for hadronization at the Relativistic Heavy Ion Collider (RHIC) have advanced our understanding of this hot and dense novel state of matter (Fig.~\ref{fig0.1}). The evolution of the system produced in the collision is explained quite successfully using the powerful method of hydrodynamics~\cite{kolb}.  However, numerous important questions like; how quickly (if at all) does the plasma thermalize, what is the life time of the system, if there is a phase transition or not, are still to be addressed.

The small size ($\sim$fm) and transient  ($\sim$fm/c) nature of the collision fireball make it extremely difficult to obtain space time information of the system. Correlation between two final state particles which is stronger for smaller separation in space time as well at smaller relative momentum, provides the most direct tool to get spatio-temporal information of the system~\cite{uli_wid}.

In 1950s', Robert Hanbury Brown and Richard Q. Twiss measured the angular size of a distant star by utilizing the method of intensity interferometry for the very first time~\cite{hbt}. Theoretically the photon bunching was first explained by Purcell~\cite{pur}, in one of the key experiments of quantum optics. In 1959 Goldhaber {\it et al.}~\cite{gold} observed an unexpected angular correlation among identical pions while studying the $\rho_0$ resonance, which was the first observation of intensity correlation in particle physics. Very soon it was realized that the correlations of identical particles emitted by highly excited sources are sensitive not only to the size of the system, but also to its lifetime and the momentum dependent correlation function contains information about the dynamics~\cite{scot1} of the system produced in the collisions.  

The heavy ion community often uses the term HBT in reference to the original work of Hanbury Brown and Twiss for any analysis  related to the spatio-temporal aspects of the particle emitting source. Several interesting theoretical works on correlation study in heavy ion collisions were reported in 1970s' by Shuryak~\cite{sur}, Gyulassy {\it et al.}~\cite{gyu} and many others. Other important contributions in the eighties include parametrization of the source function and a more detailed analysis of the role of final state interactions in the system (see review article~\cite{uli_wid} for detail).

\begin{figure}
\includegraphics[scale=0.25]{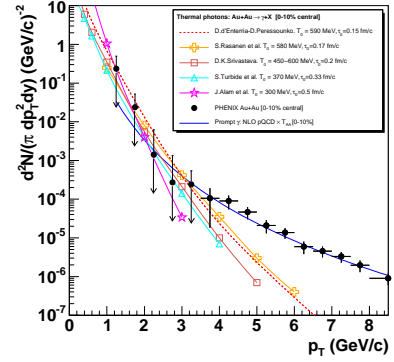}
\caption{\label{fig0.4} Thermal photon predictions for central Au+Au reactions at RHIC computed with different hydrodynamical and dynamical fireball models along PHENIX experimental data~\cite{david}.}
\end{figure}

\section{Intensity interferometry of photons} 
Electromagnetic radiation, known as the thermometer of the medium, has long been considered as one of the most promising and  efficient probes for the hot and dense state of matter produced in the relativistic collision of heavy nuclei~\cite{david}. Intensity interferometry with identical hadrons or light nuclei has been an important tool to learn about the dynamics of the subatomic or nuclear collisions. However, the hadrons suffer final state interactions and their correlations mainly carry information about the late dilute stage of the system.

Photons on the other hand, decouple from the system immediately after their creation, suffering negligible re-scatterings with the medium ($\alpha \ll \alpha_s$ ), and carry undistorted information  about the circumstances of their production to the detector. Obtaining direct information about the earliest hot and dense stage of the system by studying large $K_T$ photons is very promising. A large production of photons having very high transverse momenta is expected from the pre-equilibrium stage, and some additional sources of photons have also been proposed recently~\cite{bms_phot,fms_phot}.  The major problem of-course arises from the meager emission of direct photons compared to the huge background of decay photons (mostly from $\pi^0$ and $\eta$ mesons) produced in the collisions. Of late, there have also been tremendous advances in methods for identification of single photons~\cite{phenix}.

The theory of the intensity interferometry of photons from relativistic heavy ion collisions has been pursued in considerable detail by several authors~\cite{dks1, dks2, ors} in last two decades. These studies further refined the early expectations of photon as powerful tool to get spatio-temporal aspects of the fireball evolution. In a recent study by Frodermann and Heinz~\cite{uli}, the details of angular dependent intensity correlation of thermal photons from non-central collision of heavy nuclei as well the dual nature of thermal photon emission resulting from the superposition of QGP and hadron resonance gas photon production have been highlighted. 

It is generally felt that the experimental efforts for these studies  have a larger likelihood of success at RHIC and LHC energies because of larger initial temperature of the plasma  and a large suppression of pions due to jet-quenching.

Till date only one measurement of photon intensity correlation at very low transverse momentum $K_T$ has been reported by WA98 collaboration for central collision of lead nuclei at CERN SPS~\cite{wa98}.

\subsection{Formulation}
The spin averaged intensity correlation between two photons with momenta ${\bf k_1}$ and 
${\bf k_2}$, emitted from a completely chaotic source can be expressed in terms of the space-time emission function $S(x,{\bf K})$  as~\cite{dks1,dks3}:
\begin{equation}
C(\mathbf{q},\mathbf{K})=1+\frac{1}{2}
\frac{\left|\int \, d^4x \, S(x,\mathbf{K})
e^{ix \cdot q}\right|^2} {\int\, d^4x \,S(x,\mathbf{k_1}) \,\, \int d^4x \,
S(x,\mathbf{k_2})}
\label{def}
\end{equation}
where the obtained correlation is a function of the relative  $\mathbf{q} \, (=\mathbf{k_1}-\mathbf{k_2})$ and average momenta $\mathbf{K} \, (=(\mathbf{k_1}+\mathbf{k_2})/2)$ of the two photons. 
The space-time emission function $S$ is often approximated as the rate of production of photons, $\rm {EdN/d^4x d^3k}$ from the quark and hadronic matter phases produced in the collisions and the correlation function $C({\bf q},{\bf K})$ is decomposed in terms of the outward, sideward,  
and longitudinal momentum differences; $q_{\text{o}}$, 
$q_{\text{s}}$ and $q_{\ell}$ respectively. One can write the four-momentum of the $i$th photon in-terms of transverse momentum $k_T$, rapidity $y$, and azimuthal angle $\psi$ as,
\begin{equation}
\mathbf{k_i}=(k_{iT}\,\cos \psi_i,k_{iT} \sin \psi_i,k_{iT} \, \sinh\, y_i), 
\end{equation}
and the three momentum differences $q_{\text{o}}$, $q_{\text{s}}$ and $q_{\ell}$ are of the form:
\begin{eqnarray}
q_{\text{o}}&=&\frac{\mathbf{q_T}\cdot \mathbf{K_T}}{K_T}\nonumber\\
       &=& \frac{(k_{1T}^2-k_{2T}^2)}
         {\sqrt{k_{1T}^2+k_{2T}^2+2 k_{1T} k_{2T} \cos (\psi_1-\psi_2)}}\nonumber\\
\label{q_o}
q_{\text{s}}&=&\left|\mathbf{q_T}-q_{\text{o}}
       \frac{\mathbf{K_T}}{K_T}\right|\nonumber\\
        &=&\frac{2k_{1T}k_{2T}\sqrt{1-\cos^2(\psi_1-\psi_2)}}
      {\sqrt{k_{1T}^2+k_{2T}^2+2 k_{1T} k_{2T} \cos (\psi_1-\psi_2)}} \nonumber\\ 
\label{q_s}
q_{\ell}&=&k_{1z}-k_{2z}\nonumber\\
        &=&k_{1T} \sinh y_1 - k_{2T} \sinh y_2 \, .
\label{q_l}
\end{eqnarray}
The radii corresponding to these momentum differences are obtained by approximating the correlation function to Gaussian parametrization as;
\begin{eqnarray}
 C(q_{\text{o}}, q_{\text{s}},q_{\ell}) = 1 +  \frac{1}{2}\exp 
\left[-\left(q_{\text{o}}^2R_{\text{o}}^2 +  q_{\text{s}}^2R_{\text{s}}^2 \right. \right. \nonumber\\ 
+  \left. \left. q_{\ell}^2R_{\ell}^2 \right ) \right].
\label{c1}
\end{eqnarray}
The root mean square momentum difference $\langle 
q_i^2 \rangle$  and the radii are obtained as,
\begin{eqnarray}
\langle q_i^2 \rangle \, = \, \frac{\int \, (C-1)\, q_i^2 \, dq_i}{\int \, (C-1) \, dq_i},
\label{qi}
\end{eqnarray}
\begin{eqnarray}
R_i^2 \, = \, \frac{1}{2 \, \langle q_i^2 \rangle} \, .
\label{ri}
\end{eqnarray}
It is to be noted that the $1/[2\, \langle q_i^2 \rangle]^{1/2}$ becomes a useful measure when the 
correlation function has a more complex nature. 
\begin{figure}
\includegraphics[scale=0.4]{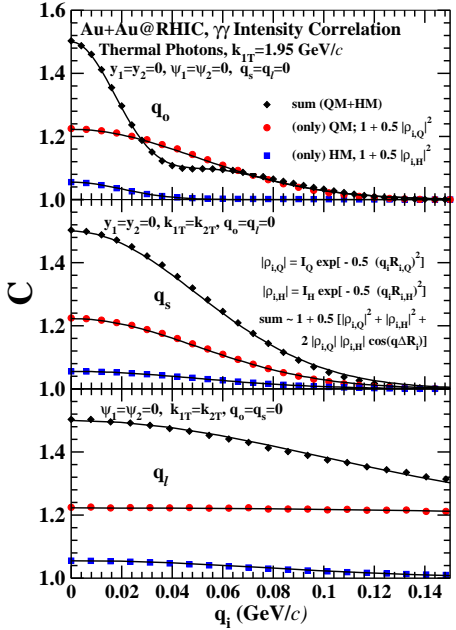}
\caption{\label{fig1} Outward (top panel), sideward (middle panel), and the longitudinal (lower panel) correlation functions for thermal photons produced in central collision of gold nuclei at RHIC. Symbols denote results of the correlation; curves denote fits~\cite{dks3}.}
\end{figure}

\subsection{Initialization and system evolution }
The exciting possibility of observation of the interference of thermal photons in the intermediate $K_T \, (\le \, 2$ GeV/$c$)  range has been explored in detail in a recent work of Srivastava {\it et al.}~\cite{dks3}. Photons in this momentum range are unique as they get competing contribution from the two phases; they have their origin either in the hot and dense quark phase of the system or in the relatively cooler but rapidly expanding hadronic phase, where a large build up of the radial flow boosts their transverse momentum. This competition gives rise to a rich structure specially in the outward correlation function, owing to the interference between the photons from the two sources. 

For this particular study, central collisions of gold and lead nuclei at the top RHIC and LHC energies respectively are considered. It is assumed that a thermally and chemically equilibrated plasma is formed at a very small initial time $\tau_0 $ (0.2 fm/$c$ for RHIC and 0.1 fm/$c$ for LHC). An isentropic expansion of the plasma is 
considered where the initial temperature $T_0$ is obtained from the relations,
\begin{equation}
\frac{2\pi^4}{45\zeta(3)} \frac{1}{A_T}\frac{dN}{dy}=4aT_0^3\tau_0 \, ,
\end{equation}
here $A_T$ is the transverse area of the system, dN/dy is the particle rapidity density, and $a=42.25\pi^2/90$ for a plasma of massless quarks and gluons. The initial energy density is estimated by combining 25\% contribution from binary collisions and 75\% from number of wounded nucleons. The bag model equation of state is considered for the system evolution where a first order phase transition takes place at a temperature of about 180 MeV and the freeeze-out temperature is taken as 100 MeV.  The relevant hydrodynamic equations are solved under the assumption of boost invariant longitudinal and azimuthally symmetric transverse expansion using the procedure discussed in Ref.~\cite{dks_hyd} and integration performed over the history of evolution. Fig.~\ref{fig0.4} shows thermal photon results from from hydrodynamic model by different groups along with direct photon data from PHENIX.

The complete leading order results for production of photons from quark matter by Arnold {\it et al.}~\cite{AMY}, and the results of Turbide {\it et al.}~\cite{TRG} for radiation from the hadronic matter are used for this analysis. The values of charge particle multiplicity at mid rapidity are taken as 1260~\cite{fms_phot} for 200A GeV Au+Au collisions at RHIC and 5625~\cite{LHC} for 5.5A TeV Pb+Pb 
collisions at LHC. 

\subsection{Correlation functions for thermal photons} 
The results for the outward, sideward, and longitudinal correlation functions for thermal photons at RHIC having $K_T \, \approx \, 2$ GeV/$c$  are shown in Fig.~\ref{fig1}. The four-momenta of the two photons are chosen so that when the outward correlation is studied, $q_s$ and $q_\ell$ are identically zero and the dependence on $q_0$ is clearly seen, and so on. Considering only the quark matter or the hadronic matter only, it is found that the correlation functions for the two phases can be approximated as
\begin{figure}
\includegraphics[scale=0.25]{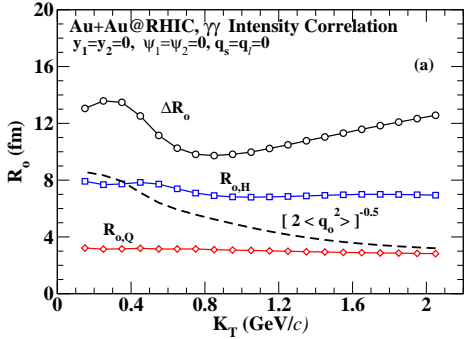}
\caption{\label{fig2} Transverse momentum dependence of the outward radii and temporal duration obtained by fitting the final correlation function for thermal photons at RHIC. Radii determined from he root-mean-square momentum difference for the correlation function are also given for comparison~\cite{dks3}.}
\end{figure}
\begin{eqnarray}
C(q_i, \alpha) = 1 \ + \ 0.5 |\rho_{i,\alpha}|^2 
\end{eqnarray}
where $i$ = $o$, $s$, or $\ell$, and $\alpha$ denotes quark matter (Q) 
and hadronic matter (H) in an obvious notation. The source distributions $|\rho_{i,\alpha}|$ is well described as a Gaussian function,
\begin{eqnarray}
 |\rho_{i,\alpha}| \ = \ I_\alpha \ \exp \left [- \ 0.5 \ 
(q_i^2R_{i,\alpha}^2) \right ] 
\end{eqnarray}
where $I_Q = dN_Q/(dN_Q+dN_H)$ and  $ I_H = dN_H/(dN_Q +dN_H)$ are the fractions of the photons from quark matter and hadronic matter respectively.
 
The final correlation functions are then approximated as,
\begin{eqnarray}
C(q_i) \ = \ 1  \ &+&  \ 0.5 \left[ \ |\rho_{i,Q}|^2 \ +  \ 
|\rho_{i,H}|^2 \right. \nonumber\\ &+& \left.
  2 |\rho_{i,Q}| |\rho_{i,H}|  \cos (q_i\Delta R_i) 
\right ] \nonumber \\
\label{cc} 
\end{eqnarray} 
where the cosine term in the above equation clearly brings out the interference between the two sources~\cite{yves}. Here $\Delta R_i$ stands for the separation of the two sources in space and time and q is the four momentum difference.
For $K_T \ \approx \ 2$ GeV/$c$, at RHIC, the value of $\Delta R_o$ is 12.3 fm obtained from fitting whereas $\Delta R_s$ and $\Delta R_\ell$ are of the order of zero. 
\begin{figure}
\includegraphics[scale=0.25]{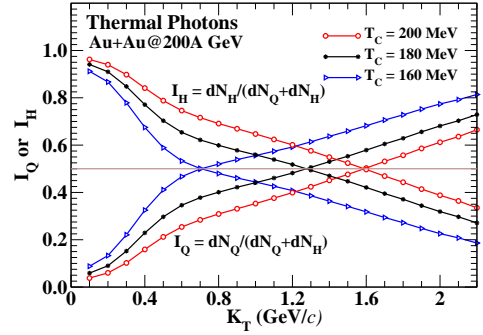}
\caption{\label{fig2.5} Transverse momentum dependence of the fraction of thermal photons from the quark matter and hadronic matter phases at RHIC with varying transition temperature~\cite{dks3}.}
\end{figure}

These results imply that, while the spatial separation of the two sources is negligible, their temporal separation is much larger, which gives the lifetime of the system. If the mixed phase is of shorter duration or absent, this will obviously decrease.

The $K_T$ dependence of $\Delta R_o$, and the outward radii for the hadronic and quark matter sources of photons obtained from fitting are shown in Fig.~\ref{fig2} which clearly explains the fact mentioned above. The outward radius for the  quark contribution depends weekly on the transverse momentum, which is indicative of a mild development of the radial flow during the quark matter phase. The corresponding radius for the hadronic contribution shows a stronger dependence on the transverse momentum resulting from a more robust development of radial flow during the late hadronic phase.

A very interesting structure is observed for the duration of the source, which increases slightly at higher $K_T$ as the momenta of the photons emitted from the hadronic phase is blue shifted to 
much larger value due to strong radial flow. The saturation of $\Delta R_o$ towards low $K_T$ has its origin in the competition between radial expansion and decoupling of hadronic matter as it cools down below the temperature of freeze-out at the edges. 

As seen from Eq.~(\ref{ri}), the inverse root mean square momentum, which is a measure of the correlation radius, is seen to vary rapidly  with transverse momentum for the outward correlation owing to rapid variation of the competing contribution of photons from the two phases. 

As the nature of the overall correlation function strongly depends on the relative contribution from the two phases, the fraction of the quark ($I_Q$) and hadronic ($I_H$) contributions are investigated  as a function of transverse momentum as well as varying critical temperature (see Fig.~\ref{fig2.5}). One can see that the transverse momentum at which the quark and hadronic contributions become equal is sensitive to the transition temperature and a decreasing $T_C$ increases the fraction of photons coming from the quark matter.
\begin{figure}
\includegraphics[scale=0.25]{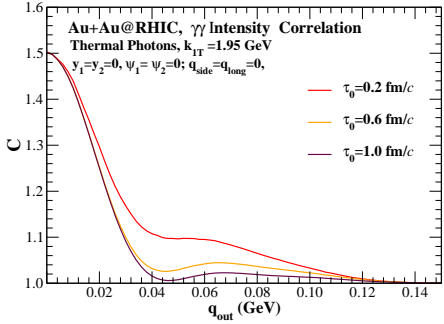}
\caption{\label{fig3} $\tau_0$ dependence of the outward correlation function at RHIC~\cite{dks3}.}
\end{figure}

The variation of the source function as a function of transverse distance and time for different values of transverse momentum has been studied in detail (not shown here) in order to understand the correlations results and the relative contributions of the two phases in it. It is observed that the radial distribution for both the phases are centered at $r_T$  near zero and the source function for the hadronic phase extends considerably beyond the same for the QGP phase. This is of course due the large transverse expansion of the system and results in a good amount of thermal photon production from hadronic phase at larger radii. A more valuable insight is provided by the temporal distribution of the source function, where a large duration of the hadronic phase is observed which controls the parameter $\Delta R_o$.

\subsection{Sensitivity to formation time and transition temperature}
We know that a smaller value of the initial formation time of the plasma implies a larger initial temperature. The sensitivity of the correlation functions to the initial formation time of the plasma is investigated by considering systems with identical entropies but varying formation time. Smaller $\tau_0$ leads to a larger production of photons from the quark matter phase and as a result the outward correlation function changes its nature as shown in Fig.~\ref{fig3}.

It is to be noted that a larger value of formation time ($\sim 1$ fm/$c$) would necessitate inclusion of the pre-equilibrium contribution of the photons, that must surely be there at least at larger values of the transverse momentum. Another important contribution, photons from jet conversion processes also likely to contribute at higher $K_T$. These contributions would increase the fraction of quark matter contribution in the correlation measurement, which in turn could mimic an effective smaller value of the formation time. 
\begin{figure}
\includegraphics[scale=0.25]{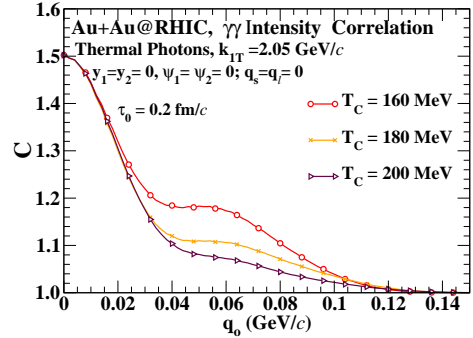}% 
\caption{\label{fig4} $T_C$ dependence of the outward correlation function at RHIC~\cite{dks3}.}
\end{figure}

In another potentially powerful observation the sensitivity of the outward correlation function to the transition temperature $T_C$ is shown in Fig.~\ref{fig4}, where the change in the interference pattern at larger $q_0$ can be easily understood in terms of the relative life time of the two phases.

The correlation results for Pb+Pb collisions at LHC energy also show similar qualitative nature as observed at RHIC. As the initial temperature likely to be attained at LHC is much larger than that at RHIC, the study of the properties of QGP phase and the dynamics of its evolution will be 
more interesting at LHC. A higher initial temperature would lead to a longer duration of the interacting system, which in turn would provide ample opportunity for the mechanism of expansion to develop.

\begin{figure}
\includegraphics[scale=0.25]{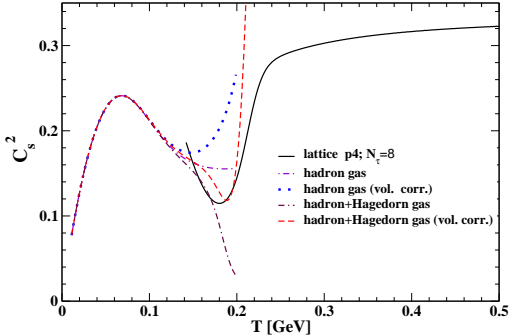}% 
\caption{\label{fig5} Speed of sound for (i) hadron resonance gas, (ii) volume corrected hadron resonance gas, (iii)hadron and Hagedorn resonance gas, and (iv) volume corrected hadron and Hagedorn resonance gas with lattice results~\cite{somnath}.}
\end{figure}

\section{Equation of state and intensity correlation of thermal photons}
In a very recent work by De {\it et al.}~\cite{somnath} the equation of state (EOS) of strongly interacting matter and how the change in EOS affects the results of  system evolution, particle spectra as well as the intensity interferometry of thermal photons have been explored in detail. It is shown that an equation of state for hot hadronic matter consisting of all baryons having $M < 2$ GeV and all mesons having $M < 1.5$ GeV, 
along with Hagedorn resonances~\cite{hag} in thermal and chemical equilibrium, matches rather smoothly with lattice equation of state (p4 action, ${N_\tau}=8$)~\cite{latt} for zero baryon chemical potential for T up to $\sim 200$ MeV, when corrected for the finite volume  of hadrons. 
\begin{figure}
\includegraphics[scale=0.25]{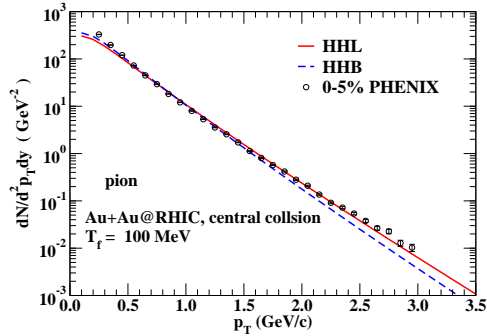}% 
\caption{\label{fig6} Pion $p_T$ spectra for the equation of state, HHB and HHL for central collisions of Au nuclei at RHIC. Experimental data for pions (0-5\%  centrality bin)~\cite{phenix_data} are shown for comparison~\cite{somnath}.}
\end{figure}

Furthermore, two equations of state for strongly interacting matter are constructed;  
one, HHL, in which the above is matched to the lattice equation of state at $T=165$ 
MeV and the other, HHB, where it is matched to a bag model equation of state with 
critical temperature $T_C=165$ MeV. The lattice equation of state displays a sharp 
cross-over for $180 \, < \,  T \, < \, 190$ MeV.

The results for the energy density, 
pressure, and the entropy density of the hadronic matter with successively increasing 
richness of the description; viz., hadron gas, hadron+Hagedorn gas, and volume 
corrected hadron+ Hagedorn gas have been studied in detail (see Ref.~\cite{somnath} for
 detail). Best agreement with the lattice calculation is obtained when the hadron + Hagedorn 
gas is corrected for the finite volume of the particles and as a result the hadronic matter for both HHL and HHB  equations of state is constructed by combining hadron and Hagedorn gas with volume correction. The results for the square of speed of sound for the four description of the hadronic matter and their comparison with the one obtained from the lattice calculations are shown in Fig.~\ref{fig5}.

\begin{figure}
\includegraphics[scale=0.25]{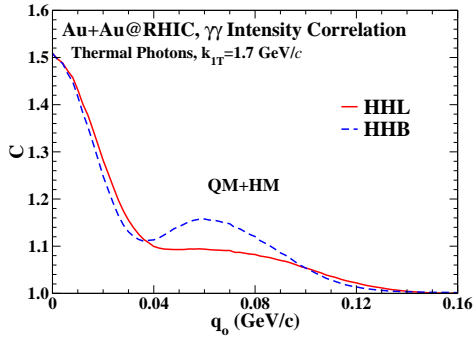}% 
\caption{\label{fig7} The outward correlation function for thermal photons at RHIC~\cite{somnath}.}
\end{figure}

\subsection{Particle spectra for HHB and HHL equations of state}
Considering an ideal hydrodynamic (azimuthally symmetric and longitudinally boost invariant) evolution of the system and the same initial conditions as used in Ref.~\cite{dks3} for Au+Au collisions at RHIC as well as for Pb+Pb collisions at LHC, transverse momentum spectra of various hadrons and photons are calculated for the two equations of state. The freeze-out  is assumed to take place at a temperature of about 100 MeV and the particle spectra are obtained using Cooper-Frye formulation~\cite{cooper}.  Whereas for the photon production, the standard rates from the Refs.~\cite{AMY,TRG} are used.
\begin{figure}
\includegraphics[scale=0.25]{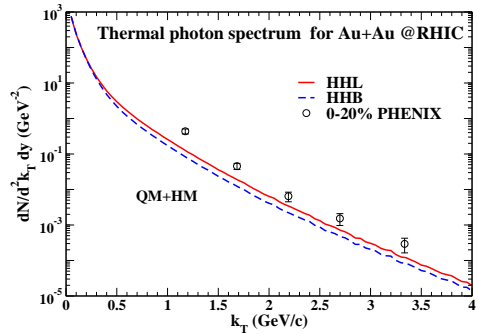}% 
\caption{\label{fig6.1} Thermal photon spectra for the equation of state, HHB and HHL for central collisions of Au nuclei at RHIC. Experimental data for (0-20\% centrality bin)~\cite{phot} are shown for comparison~\cite{somnath}.}
\end{figure}

Transverse momentum spectra of pions and thermal photons for central collision of Au nuclei at RHIC and at midrapidity are shown in Figs.~\ref{fig6} and~\ref{fig6.1} along with experimental data from PHENIX~\cite{phenix_data,phot} for the two cases. It is  observed that both the equations of state give a reasonable description of the particle distribution with a slight preference for the HHL EOS and the inverse slope of the spectra for HHL EOS is larger than the same for the HHB EOS. Similar results have been reported for the bag model and lattice EOS in Ref.~\cite{pasi} at RHIC earlier. It is to be noted that a complete description of photon data would involve addition of prompt contribution scaled by appropriate nuclear overlapping function.

Corresponding results for the LHC (see Ref.~\cite{somnath} for detail) show that the difference between the transverse momentum distribution of photons and the hadrons for the two EOS is further reduced. However, a slight increase in the inverse slope for the HHL EOS is observed as a result of the larger lifetime of the system at the LHC energy. 

One can conclude from these results that particle spectra (even for thermal photons) can not distinguish between the equations of state, one admitting a first order phase transition and the other admitting a rapid cross-over as suggested by lattice calculations. However, it is to be noted that the history of evolution of the two systems is different as they are subjected to different rates of expansion due to the varying speed of sound.

In the same study it is observed that the results for the temporal evolution of average energy density, temperature and radial flow velocity for both the RHIC and LHC energies do not show significant difference for the two EOS HHL and HHB. The most profound variation is observed for the radial velocity for the two EOS, where the radial velocity for HHL rises continuously, stays below that for the HHB, but overshoots it once the later stalls due to the onset of the mixed phase. As a result, final $v_t$ for the HHL EOS is slightly larger than that for the HHB.

\subsection{EOS and intensity interferometry of thermal photons}
It is observed that the HHL equation state leads to a larger production of photons at smaller radial distances as well as at intermediate times. We know that the photons from quark matter originate at early times and those from hadronic matter are emitted from the later part of the system evolution As the interference between the two depends on their relative contributions, this difference holds out a promise that we may see a difference in the intensity interferometry of photons for the two equations of state.

\begin{figure}
\includegraphics[scale=0.25]{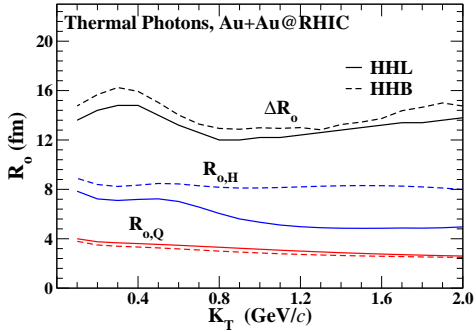}% 
\caption{\label{fig8} Transverse momentum dependence of the outward radii for hadronic and quark matter sources for thermal photons~\cite{somnath}.}
\end{figure}

It is found that the sideward correlation is essentially identical for the two equations of state at both RHIC and LHC energies, whereas the longitudinal correlation function shows a slight variation when the EOS is changed from HHB to HHL (see Ref.~\cite{somnath} for details). The results for the outward correlation (shown in Fig.~\ref{fig7}) are more dramatic and show a clear difference for the two equations of state. The same treatment of Ref.~\cite{dks3} is followed to estimate the source size of the hadronic and the quark matter contributions. 

A very clear difference in the
underlying radii for the hadronic matter contribution (shown in Fig.~\ref{fig8}) for the two equations of state is observed; where the one for the HHB equation of state being effectively much larger. 

One can see that the difference $\Delta R_o$, which corresponds to the lifetime of the system is slightly smaller for the HHL equation of state, as it does not incorporate mixed phase like the HHB. The $R_o$ for the hadronic matter for the HHL equation of state is also much smaller for the same reason. 

Similar qualitative nature for the correlation functions and radii is observed at LHC energy also. In view of these important observations, it would be quite interesting to analyze this behavior using a fully 3+1D hydrodynamics~\cite{nonaka}.

\section{Summary and Conclusions}
The interference of thermal photons from the quark matter and hadronic matter phases produced in relativistic heavy ion collisions gives rise to a unique structure, specially in the outward intensity correlation function for intermediate transverse momenta. The correlation results are found to be quite sensitive to the initial conditions and the equation of state of the strongly interacting matter. This can be used to distinguish between the bag model and lattice based equations of state.

\section{Acknowledgments}
RC and SD gratefully acknowledge the financial supports from  Finnish Academy project 'Hard processes in heavy-ion 
collisions at the Large Hadron Collider' and the  Department of Atomic Energy respectively.

%\section*{Acknowledgments}
%\acknowledgments

%\end{acknowledgments}

%\ \\
%\noindent


\begin{thebibliography}{50}


\bibitem{fl1} C.~Adler {\it et al.} [STAR Collaboration], Phys.\ Rev. 
\ Lett. {\bf 87}, 182301 (2001); ibid {\bf 89}, 132301 (2002); ibid 
{\bf 90}, 032301 (2003); S.~S.~Adler {\it et al.} [PHENIX Collaboration], 
Phys.\ Rev. \ Lett. {\bf 91}, 182301 (2003)


\bibitem{fl2} P.~Huovinen, P.~Kolb, U.~Heinz, P.~V.~Ruuskanen, and
S.~Voloshin, \ Phys.\ Lett. \ B {\bf 503}, 58 (2001); D.~Teaney, J.~Lauret,
 and E.~Shuryak, nucl-th/0110037.

\bibitem{jet1} X.~N.~Wang, Phys.\ Rev. \ C {\bf 63}, 054902 (2001);
M.~Gyulassy, I.~Vitev, and X.~N.~Wang, Phys.\ Rev. \ Lett. {\bf 86},
 2537 (2001).

\bibitem{jet2} K.~Adcox {\it et al.} [PHENIX Collaboration], Phys.
\ Rev.\ Lett. {\bf 88}, 022301 (2002); J.~Adams {\it et al.} 
[STAR Collaboration], Phys.\  Rev.\ Lett. {\bf 91}, 172302 (2003). 


\bibitem{rec} V.~Greco, C.~M.~Ko and P.~Levai, \ Phys.\ Rev.\ Lett. 
{\bf 90} 202302 (2003); R.~J.~Fries, B.~M\"uller, C.~Nonaka, and S.~A.~Bass,
\ Phys. \ Rev.\ Lett. {\bf 90}, 202303 (2003).

\bibitem{kolb}
  P.~Kolb and U.~Heinz,
  ``Hydrodynamic description of ultrarelativistic heavy-ion collisions,''
  arXiv:nucl-th/0305084.


\bibitem{uli_wid} U. A. Wiedemann and U. Heinz, Phys. \ Rept. \ {\bf 319}, 145 (1999).

\bibitem{hbt} R. H. Brown and R. Q. Twiss, \ Nature {\bf 177}, 27 (1956); {\it ibid.} {\bf 178}, 1046 
(1956); \ Proc. \ Roy. \ Soc. {\bf A242}, 300 (1957); {\it ibid.} {\bf 243}, 291 (1957).


\bibitem{pur} E. Purcell, \ Nature {\bf 178}, 1449 (1956).



\bibitem{gold} G. Goldhaber, S. Goldhaber, W. Lee and A. Pais, Phys. \ Rev. {\bf 120}, 300
(1960).

\bibitem{scot1}S.~Pratt,
  Phys.\ Rev.\  D {\bf 33}, 1314 (1986), G.~Bertsch, M.~Gong and M.~Tohyama,
  Phys.\ Rev.\  C {\bf 37}, 1896 (1988),G.~F.~Bertsch,
  Nucl.\ Phys.\  A {\bf 498}, 173C (1989).

\bibitem{sur} E. Shuryak, Phys. \ Lett. {\bf B44}, 387 (1973); Sov. \ J. \ Nucl. \ Phys. {\bf 18}, 667
(1974).


\bibitem{gyu} M. Gyulassy, S.K. Kauffmann, and L.W. Wilson, Phys. \ Rev. \ C {\bf 20}, 2267
(1979).

\bibitem{flf} J. Adams {\it et al.} [STAR Collaboration] \ Phys. \ Rev. \ Lett. {\bf 92}, 052302 (2004).

\bibitem{jetf} S. S. Adler {\it et al.} [PHENIX Collaboration], \ Phys. \ Rev. \ Lett. {\bf 96}, 202301 (2006).

\bibitem{recof} A. Adare {\it et al.} [PHENIX Collaboration], \ Phys. \ Rev. \ Lett. {\bf 98}, 162301 (2007).

\bibitem{david} D. G. d'Enterria and D. Peressounko, \ Eur. \ Phys. \ J. \ C 
{\bf 46}, 451 (2006).

\bibitem{bms_phot} S.~A.~Bass, B.~M\"uller, and D.~K.~Srivastava,
\ Phys. \ Rev. \ Lett. {\bf 90}, 082301 (2003).




\bibitem{fms_phot} R.~J.~Fries, B.~M\"uller, and D.~K.~Srivastava,
\ Phys. \ Rev. \ Lett. {\bf 90}, 132301 (2003).

\bibitem{phenix}  T. Dahms [PHENIX Collaboration], \ J. \ Phys. 
\ G {\bf 35}, 104118 (2008); A. Adare { \it et al.} 
[PHENIX Collaboration], arXiv:0804.4168 [nucl-ex]. 

\bibitem{dks1}D.~K.~Srivastava,  \ Phys. \ Rev. \ C {\bf 71}, 034905 (2005).

\bibitem{dks2}D.~K.~Srivastava and J.~Kapusta,  Phys.\ Lett.
\ B {\bf 307}, 1 (1993); D.~K.~Srivastava and J.~Kapusta,  Phys. 
\ Rev.\ C {\bf 48}, 1335 (1993); D.~K.~Srivastava, \ Phys.  Rev. 
\ D {\bf 49}, 4523 (1994); D.~K.~Srivastava and C.~Gale,  Phys. 
\ Lett. \ B{\bf 319}, 407 (1993); D.~K.~Srivastava and J.~Kapusta, 
 Phys.\ Rev.\ C {\bf 50}, 505 (1994); S.~A.~Bass, B.~M\"uller, 
and D.~K.~Srivastava, Phys. \ Rev. \ Lett. {\bf 93}, 162301 (2004).

\bibitem{ors} A.~Timmermann, M.~Pl\"umer, L.~Razumov, and R.~M.~ Weiner,
  Phys.\  Rev. \ C {\bf 50}, 3060 (1994); J.~Pisut, N.~Pisutova, and 
B.~Tomasik,  Phys.\ Lett.\ B {\bf 345}, 553 (1995); C.~Slotta and U.~Heinz,  Phys.\ Lett.\ B 
{\bf 391}, 469 (1997); D.~Peressounko,  Phys. \ Rev. \ C {\bf 67}, 
014905 (2003); J.~Alam {\it et al.},  Phys. \ Rev. \ C {\bf 67}, 054902 (2003);  T.~Renk, hep-ph/0408218.

\bibitem{uli}E.~Frodermann and U.~Heinz,
  Phys.\ Rev.\  C {\bf 80}, 044903 (2009).


\bibitem{wa98} M.~M.~Aggarwal {\it et al.} [WA98 Collaboration],  Phys. \  Rev. \ Lett. {\bf 93}, 022301 (2004).




\bibitem{dks3}
  D.~K.~Srivastava and R.~Chatterjee,
  Phys.\ Rev.\  C {\bf 80}, 054914 (2009) 
  [Erratum-ibid.\  C {\bf 81}, 029901 (2010)].
%\bibitem{wa98} M.~M.~Aggarwal {\it et al.} [WA98 Collaboration], \ Phys. \ Rev. \ Lett. {\bf 85}, 3595 (2000).

\bibitem{dks_hyd}H.~Von Gersdorff, L.~D.~McLerran, M.~Kataja and
 P.~V.~Ruuskanen, Phys.\ Rev.\  D {\bf 34} (1986) 794,
J.~Cleymans, K.~Redlich, and D.~K.~Srivastava, 
\ Phys. \ Rev. \ C {\bf 55}, 1431 (1997).
\bibitem{AMY} P.~Arnold, G.~D.~Moore, and L.~G.~Yaffe, JHEP {\bf 12},
009 (2001).

\bibitem{TRG} S.~Turbide, R.~Rapp, and C.~Gale,
 \ Phys.\ Rev. \ C {\bf 69}, 014903 (2004).

\bibitem{LHC} J.~Kapusta, L.~McLerran, and D.~K.~Srivastava, \ Phys. \ Lett.
B {\bf 283}, 145 (1992).

\bibitem{yves} F.~Marques, G. Martinez, G. Matulewicz, R. Ostendorf, and Y. Schutz, \ Phys. \ Rept. {\bf 284}, 91 (1997); F. Marques {\it et al.}, Phys. \ Rev.\ Lett. {\bf 73}, 34 (1994); Phys. \ Lett. {\bf B349}, 30 (1995); {\bf B394}, 37 (1997).

\bibitem{somnath} S. De, D. K. Srivastava, and R. Chatterjee, \ J. \ Phys. \ G {\bf 37}, 115004 (2010). 

\bibitem{hag}
  R.~Hagedorn,
  Nuovo Cim.\ Suppl.\  {\bf 3} (1965) 147.

\bibitem{latt} 
  A.~Bazavov {\it et al.},
  Phys.\ Rev.\  D {\bf 80}, 014504 (2009).

\bibitem{cooper}F.~Cooper and G.~Frye, Phys. \ Rev. \ D {\bf 10}, 186 (1974).


\bibitem{phenix_data}S.~S.~Adler {\it et al.}  [PHENIX Collaboration],
  Phys.\ Rev.\  C {\bf 69}, 034909 (2004).

\bibitem{phot}
  A.~Adare {\it et al.}  [PHENIX Collaboration],
  Phys.\ Rev.\ Lett.\  {\bf 104}, 132301 (2010).

\bibitem{pasi} P.~Huovinen, Nucl. \ Phys. \ A {\bf 761}, 296 (2005),
 P.~Huovinen and P.~Petreczky,  Nucl.\ Phys.\  A {\bf 837}, 26 (2010).


\bibitem{nonaka} See e.g., C.~Nonaka and S.~A.~Bass,
  Phys.\ Rev.\  C {\bf 75}, 014902 (2007).

\end{thebibliography}
\end{document}